\documentclass[12pt,a4paper]{article}

\newcommand{\p}{\partial}
\newcommand{\la}{\lambda}
\newcommand{\e}{\epsilon}

\newcommand{\ph}{\varphi}
\newcommand{\dA}{\dot{A}}
\newcommand{\dB}{\dot{B}}
\newcommand{\dC}{\dot{C}}
\newcommand{\dD}{\dot{D}}
\newcommand{\X}{\mathcal{X}}
\newcommand{\Y}{\mathcal{Y}}
\newcommand{\1}{\hat{1}}
\newcommand{\2}{\hat{2}}
\newcommand{\3}{\hat{3}}

\newcommand{\ba}{\begin{align}}
\newcommand{\bpm}{\begin{pmatrix}}
\newcommand{\epm}{\end{pmatrix}}
\newcommand{\be}{\begin{equation}}
\newcommand{\ee}{\end{equation}}

\usepackage{amsmath,amsthm,amsfonts}

\begin{document}

\centerline{\Large \bf{Towards conformally flat isothermic metrics}}

\bigskip

\centerline{A. Szereszewski$^\dag$ and A. Sym$^\ddag$}

\bigskip
\noindent
$^\dag$ Institute of Theoretical Physics, Faculty of Physics, University of Warsaw,  Poland\\
$^\ddag$ Department of Mathematical Methods in Physics, Faculty of Physics, University of Warsaw,  Poland

\bigskip
\noindent
E-mail: aszer@fuw.edu.pl

\begin{abstract}
According to \cite{ASAS} if the stationary Schr\"odinger equation on $n$-dim. Riemann space
admits $R$-separation of variables (i.e. separation of variables with a
factor $R$), then the underlying metric is necessarily isothermic. An
important sub-class of isothermic metrics are the so called binary
metrics. In this paper we study  conditions for vanishing of components
$C_{ijkl}$ of Weyl tensor of arbitrary 4-binary metrics.  In particular all 
4-binary metrics for which $C_{ijij}$ are the only non-vanishing components
are classified into four classes. Finally, conformally flat metrics of the
last class are isolated.
\end{abstract}

\section{Introduction}
   
The method of separation of variables is one of the most useful method of solving  linear partial 
differential equations. The theory of $R$-separability of variables (separability with a factor $R(x)$) 
in the 3-dimensional Laplace equation were laid in the second part of nineteenth century with the work of 
G. Darboux \cite{D7}. The detailed study of
orthogonal coordinates in Euclidean spaces has also been given by him in \cite{D6} and \cite{D8}. 
The contributions to separability of Schr\"odinger equation in $n$-dimensional Euclidean space were made 
by H.P. Robertson \cite{HR} and L.P. Eisenhart \cite{LE}. In twentieth century W. Miller and
E.G. Kalnins studied separability theory also in connection with symmetries (see e.g. \cite{WM2}). 

It is also of great importance to find and analyze exact solutions on non-flat spaces.   
The method of separation allows to construct such solutions and moreover  
imposes the restrictions on the metrics. 

This article is a part of the project of describing 
the geometry of isothermic metrics.


\subsection{$R$-separability}

We assume that 4-dimensional  space $\mathcal{R}^4$ admits local orthogonal coordinates $x=(x^1,x^2,x^3,x^4)$ 
in which the metric has the following form
 \be
   g=\sum_{i=1}^4 H_i^2(dx^i)^2, \qquad H_i=H_i(x). \label{diag_met}
 \ee
In  \cite{ASAS} we investigated the problem of $R$-separability of $n$-dimensional  stationary Schr\"odinger 
equation
 \be
 \Delta\psi+(k^2-V(x))\psi=0,\qquad\Delta= h^{-1}\sum_{i=1}^n\frac{\p}{\p x^i}\frac{h}{H_i^2}\frac{\p}{\p x^i},
\qquad h=\prod_{i=1}^nH_i,\qquad k=const,  \label{Schr}   
 \ee
i.e. the existence of solution of the form $\psi(x)=R(x)\prod_i\psi_i(x^i)$, where $R(x)$ is a non-vanishing 
function and functions of one variable
$\psi_i(x^i)$ satisfy 
 \be 
     \psi''_i+p_i\psi'_i+q_i\psi_i=0 
 \ee 
for some $p_i(x^i),q_i(x^i)$. In has also been shown that the stationary Schr\"odinger equation (\ref{Schr})
is $R$-separable in $\mathcal{R}^n$ if the metric is isothermic and the $R$-equation is satisfied 
(for details see Theorem 3 in \cite{ASAS}). 

The isothermic 4-dimensional metric is given by (\ref{diag_met}) with 
 \be
  H_i^2=R^{-2}G_{(i)}^{-2}f_i^{-2}\prod_{k=1}^4 G_{(k)},  \label{isoth_H_i}
 \ee
where  $G_{(i)}$  does not depend on  $x^i$  while  $f_i$ depends only on $x^i$.
Given six functions $G_{ij}(x^i, x^j)$  $(i,j = 1,2,3,4$ and $i<j)$. In (\ref{isoth_H_i}) we
put
 \be
     G_{(i)} = \prod_{p,q\neq i} G_{pq} 
 \ee 
then $H_i$ is given explicitly as
  \begin{equation}
   H_1=\frac{G_{12} G_{13} G_{14}}{Mf_1}, \quad H_2=\frac{G_{12} G_{23} G_{24}}{Mf_2}, \quad
   H_3=\frac{G_{13} G_{23} G_{34}}{Mf_3}, \quad H_4=\frac{G_{14} G_{24} G_{34}}{Mf_4},  
   \label{H_bin_met}
 \end{equation}
where $G_{ij}$ depend only on two variables $x^i$ and $x^j$ and $M$ stands for $R$.
The metric (\ref{diag_met}) where $H_i$ are given by  (\ref{H_bin_met}) is called binary metric.




\subsection{Summary convention}

It is difficult to maintain the summation convention when working with
diagonal metrics. Therefore Einstein summary convention is not used throughout
this paper.


\section{Geometric quantities for binary metrics}

 We will investigate the binary metric (\ref{diag_met}), (\ref{H_bin_met}) with specific Weyl 
tensor. We introduce new functions $\ph_{ij}(x^i,x^j)$ defined by
 \[ G_{ij}=\exp(\ph_{ij}) \]   
and use them to rewrite binary metric in the form
 \be
   g=\frac{1}{M^2}\left[e^{\ph_{12}+\ph_{13}+\ph_{14}}\frac{(dx^1)^2}{F_1}+
                        e^{\ph_{12}+\ph_{23}+\ph_{24}}\frac{(dx^2)^2}{F_2}+
                        e^{\ph_{13}+\ph_{23}+\ph_{34}}\frac{(dx^3)^2}{F_3}+
                        e^{\ph_{14}+\ph_{24}+\ph_{34}}\frac{(dx^4)^2}{F_4}\right],   \label{bin_metr}
  \ee
where $F_i=f_i^2(x^i)$. From now on we assume that $F_i$ are arbitrary functions which could be also negative.
The Weyl tensor of binary metric (\ref{bin_metr}) is given by (all indicies $i,j,k,l$ are different
and range from 1 to 4)
 \begin{align}
  C^i\,_{jkl} =& 0, \qquad {i,j,k,l}-\text{different},   \label{Cijkl}\\
  C^k\,_{ikj} =& \ph_{ij,i}\ph_{jk,j}+\ph_{ki,i}\ph_{ij,j}-\ph_{ki,i}\ph_{kj,j}-
    \frac{1}{2}\sum_{l\neq
      i,j}(\ph_{ij,i}\ph_{jl,j}+\ph_{li,i}\ph_{ij,j}-\ph_{li,i}\ph_{lj,j}),\quad
         {i,j,k}-\text{different},  \label{Ckikj}  \\
  C^{ij}\,_{ij} =&-\frac{1}{3}M^2\left[\frac{1}{2}G_i^{-2}E_{ij,i}F'_i+\frac{1}{2}G_j^{-2}E_{ji,j}F'_j 
             -\frac{1}{2}\sum_{k\neq i,j}G_k^{-2}\p_k\left(E_{ki}+E_{kj}\right) F'_k \right.\nonumber\\
    &+\left. G_i^{-2}\left(E_{ij,ii}-\ph_{ij,i}\sum_{k\neq i,j}\ph_{ik,i}+
       \sum_{l\neq i,j}\sum_{k\neq i,j,l}\ph_{il,i}\ph_{ik,i}  \right)F_i\right. \nonumber\\
    &+\left. G_j^{-2}\left(E_{ji,jj}-\ph_{ij,j}\sum_{k\neq i,j}\ph_{jk,j}+
       \sum_{l\neq i,j}\sum_{k\neq i,j,l}\ph_{jl,j}\ph_{jk,j}  \right)F_j\right. \\
    & -\left. \sum_{k\neq i,j}G_k^{-2}\left(\p_k^2\left(E_{ki}+E_{kj}\right)-
     3\ph_{ik,k}\ph_{jk,k}+\frac{1}{2}\sum_{l\neq k}\sum_{r\neq k}\ph_{kl,k}\ph_{kr,k}\right)F_k\right]
       \qquad   i\neq j, \nonumber
 \end{align}
where 
 \[ E_{ij}=\ph_{ij}-\frac{1}{2}\sum_{k\neq i,j}\ph_{ik}. \]


\section{Special binary metrics}

As can be seen from (\ref{Cijkl}), for any binary metric the components of Weyl tensor 
with all indices different are zero (in fact, this is true for any diagonal metric). 
We classify all  metrics (\ref{bin_metr}) for which in addition 
 \begin{equation}
    C^k\,_{ikj}=0.   \label{Ckikj0}
 \end{equation}
If conditions (\ref{Ckikj0}) hold then there exists at most two non-zero independent components of Weyl 
tensor  (eg. $C_{1212}$, $C_{1313}$). 
To find all metrics for which  (\ref{Ckikj0}) holds we consider the derivaties of
components  of the Weyl tensor (\ref{Ckikj}), namely
 \begin{equation}
  C^l\,_{ilj,k} = - C^k\,_{ikj,k}=
  -\frac{1}{2}\big((\ph_{ij}-\ph_{ki})_{,i}\ph_{jk,jk}+(\ph_{ij}-\ph_{kj})_{,j}\ph_{ik,ik}\big), \qquad   k\neq i,j,l.
 \end{equation}
Introducing the following quantities
 \begin{equation}
  \la_{ijk}=-\la_{jik}=(\ph_{jk}-\ph_{ik})_{,k}\ph_{ij,ij}
 \end{equation}
we can rewrite $C^k\,_{ikj,k}=\la_{k(ij)}$. Therefore, the necessary condition
for (\ref{Ckikj0}) to be satisfied is 
 \be
   \la_{(ijk)}=0 , \qquad  {i,j,k}-\text{different}.    \label{lam(ijk)}
 \ee
Equations (\ref{lam(ijk)}) can be rewritten as
 \be
     \la_{ijk}=\la_{jki}=\la_{kij}    \label{lamijk}
 \ee
for any three different indices $i,j,k=1,2,3,4$. Surprisingly all
solutions of (\ref{lamijk}) are known \cite{LE}. They are listed in Table 1,
where funcions $U_i$, $V_i$ and $Q_i$ depend only on $x^i$ and $m$ is constant.

 \begin{table}[ht]
 \begin{center}
 \begin{tabular}{l|c|c|c|c|c|c}
     & $\ph_{12}$ &  $\ph_{13}$ &  $\ph_{14}$  & $\ph_{23}$ & $\ph_{24}$ & $\ph_{34}$ \\ \hline
   i)& arbitrary  &  $U_1+U_3$  &  $U_1+U_4$   & $U_2+U_3$  & $U_2+U_4$  & arbitrary\\ \hline
  ii)& arbitrary  &  $U_1+U_3$  &  $V_1+V_4$   & $U_2+U_3$  & $V_2+V_4$  & $Q_3+Q_4$\\ \hline  
 iii)& $U_1+U_2$  &  $V_1+U_3$  &  $Q_1+U_4$   & $V_2+V_3$  & $Q_2+V_4$  & $Q_3+Q_4$\\ \hline  
  iv)& $m\ln|U_1-U_2|$ & $m\ln|U_1-U_3|$ & $m\ln|U_1-U_4|$ &
  $m\ln|U_2-U_3|$ & $m\ln|U_2-U_4|$ & $m\ln|U_3-U_4|$\\ \hline
 \end{tabular} 
 \caption{Components of the binary metrics (\ref{bin_metr}) for which $C^k\,_{ikj}=0$.}
 \label{tab}
 \end{center}
 \end{table}

It is noted that the Weyl tensor of Lorentzian binary metrics for which (\ref{Ckikj0}) holds might 
be algebraically special of Petrov type $D$ or $O$.   

In the next subsection we concentrate on the case iv) where $m$ is integer or half integer.


\subsection{Case iv)}
In the case iv) we assume that $U_i$ ($i=1,2,3,4$) are non-constant function and  choose $U_i=x^i$.
Then the metric (\ref{bin_metr}) becomes
 \begin{multline}
   g=\frac{1}{M^2}\left[\frac{(x^1-x^2)^{2m}(x^1-x^3)^{2m}(x^1-x^4)^{2m}}{F_1(x^1)}(dx^1)^2+
                        \frac{(x^1-x^2)^{2m}(x^2-x^3)^{2m}(x^2-x^4)^{2m}}{F_2(x^2)}(dx^2)^2 \right.\\
                  \left.+\frac{(x^1-x^3)^{2m}(x^2-x^3)^{2m}(x^3-x^4)^{2m}}{F_3(x^3)}(dx^3)^2
                        +\frac{(x^1-x^4)^{2m}(x^2-x^4)^{2m}(x^3-x^4)^{2m}}{F_4(x^4)}(dx^4)^2 \right]. \label{met_iv}
 \end{multline}
We do not assume that functions $F_i$ are positive, therefore in general the latter can have Riemannian, Lorentzian
or neutral $(++--)$ signature in some regions. 
All conformally flat metrics (\ref{met_iv}) with $m=\frac{N}{2}$ ($N\in\mathbb{Z}$) can be found.\\

\noindent 
{\bf Lemma.} The metric (\ref{met_iv}) with integer or half-integer $m$ is conformally flat if 
and only if 
\begin{enumerate}
 \item[a)] $m=-1$ and $F_i(x^i)$ are constant functions,
 \item[b)] $m=-\frac{1}{2}$ and $F_i(x^i)$ are polynomials of second order,
 \item[c)] $m=0$ and $F_i(x^i)$ are arbitrary functions of one variable, 
 \item[d)] $m=\frac{1}{2}$ and  $F_i(x^i)$ are polynomials of sixth order.
\end{enumerate}

\emph{Proof:} The condition (\ref{Ckikj0}) is automatically satisfied for all metrics (\ref{met_iv}). 
Hence there exist only 2 independent components of Weyl tensor, e.g. $C^{12}\,_{12}$, 
$C^{13}\,_{13}$, which depend lineary on $F_i$ and $F'_i$.  
It turns out that 
 \ba
   &\left[x^2(x^1+x^3-2x^4)+x^3(x^4-2x^1)+x^1x^4\right)]C^{12}\,_{12}-
   \left[x^3(x^1+x^2-2x^4)+x^2(x^4-2x^1)+x^1x^4\right]C^{13}\,_{13} \nonumber\\
   &=\frac{M^2}{2}m(2m-1)\prod_{k<l}(x^k-x^l)^{-2m-1}\sum_{i=1}^4\Big(
     \prod_{\substack{k<l\\k,l\neq i}}(x^k-x^l)^{2(m+1)}\Big)F_i.
 \end{align}
There are two special cases: $m=0$ and $m=\frac{1}{2}$ which must be treated separately. 
If $m=0$ then metric (\ref{met_iv}) is obviously conformally flat for any functions $F_i(x^i)$,
 \be
   g=\frac{1}{M^2}\left[\frac{(dx^1)^2}{F_1(x^1)}+\frac{(dx^2)^2}{F_2(x^2)}+
                        \frac{(dx^3)^2}{F_3(x^3)}+\frac{(dx^4)^2}{F_4(x^4)}\right].  \label{met0}
 \ee 
In the case $m=\frac{1}{2}$ it can be shown that
 \begin{equation}
    \p_i^7F_i = 0, \qquad i=1,2,3,4,
 \end{equation}
so $F_i(x^i)$ are polynomials at most of the sixth degree.
Further analysis of the components $C^{12}\,_{12}$, $C^{13}\,_{13}$ leads to conclusion that
the coefficients of the polynomials are equal up to sign. Finally conformally flat metric in
this case can be written in the following form
 \begin{equation}
 g=\frac{1}{M^2}\sum_{i=1}^4\frac{\prod_{j\neq i}(x^i-x^j)}{\sum_{k=0}^6 a_k(x^i)^k}(dx^i)^2, \label{met1/2}
 \end{equation}
where $a_k$ are arbitrary constants. 

If $m\notin\{0,\frac{1}{2}\}$ then 
 \begin{equation}
  L:=\sum_{i=1}^4\Big(\prod_{\substack{k<l\\k,l\neq i}}(x^k-x^l)^{2(m+1)}\Big)F_i=0. \label{L=0}
 \end{equation}
By differentiate the latter with respect $x^1,x^2,x^3$ and $x^4$ we can express each $F'_i$ in terms
of $F_1,F_2,F_3, F_4$ and then eliminate $F'_i$ from  $C^{12}\,_{12}$ and $C^{13}\,_{13}$. The calculation
ends up with the result
 \begin{equation}
   C^{12}\,_{12}={\sf f_1}\,L, \qquad C^{13}\,_{13}={\sf f_2}\,L,
 \end{equation}
where $\sf{f_1},\sf{f_2}$ are two non-vanishing functions. This shows that (\ref{L=0}) is a necessary and 
sufficient condition for metric  (\ref{met_iv}) to be conformally flat. We show below that there exist
non-trivial solutions for (\ref{L=0}) only if $m=-1$ or $m=-\frac{1}{2}$ (we omit cases $m=0,\frac{1}{2}$).  

First let us assume that $m<-1$, then $\alpha=-2(m+1)>0$ and it is more convenient to use equation $K=0$, where
 \begin{multline}
    K=L\prod_{k<l}(x^k-x^l)^{-2(m+1)}\\
     =[(x^1-x^2)(x^1-x^3)(x^1-x^4)]^{-2(m+1)}F_1+[(x^1-x^2)(x^2-x^3)(x^2-x^4)]^{-2(m+1)}F_2\\
      +[(x^1-x^3)(x^2-x^3)(x^3-x^4)]^{-2(m+1)}F_3+[(x^1-x^4)(x^2-x^4)(x^3-x^4)]^{-2(m+1)}F_4.
 \end{multline} 
We can differentiate $\alpha$ times equation $K=0$ with respect to $x^1,x^2,x^3$ and $x^4$ and obtain
 \begin{equation}
   \alpha!^3\left[(-1)^\alpha F_1^{(\alpha)}+F_2^{(\alpha)}+(-1)^\alpha F_3^{(\alpha)}+F_4^{(\alpha)}\right]=0,
 \end{equation} 
where $(\alpha)$ in subscript denotes derivative of order $\alpha$.
This shows that $F_i$ must be polynomials at most of the degree $\alpha$. We proceed further with
$F_1$.
Calculating the derivative  $\p_1^{\alpha+1}\p_2^{\alpha-1}\p_3^\alpha\p_4^\alpha K$ one gets
 \begin{equation}
   (x^1-x^2)F_1^{(\alpha+1)}+(\alpha+1)F_1^{(\alpha)}=0
 \end{equation}
and so, $F_1$ is a polynomial at most of the degree $\alpha-1$. Considering now 
$\p_1^{\alpha+1}\p_2^{\alpha-l}\p_3^\alpha\p_4^\alpha K$, where $l=1,2,\dots,\alpha$ we
conclude by induction that $F_1=const$. An analogous treatment with $F_2,F_3,F_4$ leads to the 
result 
 \begin{equation} 
    F_i=const,\qquad i=1,2,3,4.
 \end{equation}
But then 
 \begin{equation}
  K=(x^1)^{-6(m+1)}F_1+(-x^2)^{-6(m+1)}F_2+(x^3)^{-6(m+1)}F_3+(-x^4)^{-6(m+1)}F_4+\mathcal{O},
   \quad F_i=const,
 \end{equation}
where $\mathcal{O}$ contains terms with $x^i$ to power less then $-6(m+1)$. We obtain from 
this $F_i=0$, $i=1,2,3,4$. This means that no solutions exist for $m<-1$.

If $m=-1$ then from (\ref{L=0}) 
 \be
   \sum_{i=1}^4 F_i=0. \label{m=0}
 \ee
Hence, the following metric
 \be
   g=\frac{1}{M^2}\sum_{i=1}^4\frac{(dx^i)^2}{F_i\prod_{j\neq i}(x^i-x^j)^2}, \label{met-1}
 \end{equation}
is conformally flat provided $F_i$ are constants and (\ref{m=0}) is satisfied. It is easily seen
that the latter can have  Riemannian, Lorentzian or neutral $(++--)$ signature. 

Let us assume that $m>-1$ and consider the identity (\ref{L=0}). Calculating derivatives of $L=0$ with respect 
to all $x^i$ we obtain four new equations which must be satisfed by $F_i$ and $F'_i$, namely
 \be
      \p_i L = 0, \qquad i=1,2,3,4.
 \ee
Now we solve the latter for $F'_i$ and insert them into
 \be
    \p_{12}L=0, \qquad   \p_{13}L=0, \qquad   \p_{14}L=0.  \label{der2_L}
 \ee
By this procedure we get four homogeneous equations involving only $F_i$ ((\ref{L=0}) and (\ref{der2_L})).
Non-zero solution is possible if and only if $\det M = 0$, where $M$ is ($4\times 4$) matrix containing
coefficients of $F_i$ ($i=1,2,3,4$) in (\ref{L=0}) and (\ref{der2_L}). It can be shown that 
 \be
  \det M = -16(m+1)^3(2m+1)^3(x^2-x^3)^2(x^2-x^4)^2(x^3-x^4)^2\prod_{k<l}(x^k-x^l)^{2(2m+1)}
 \ee
and hence $m=-1$ (which was considered before) or $m=-\frac{1}{2}$. Further analysis of the components 
$C^{12}\,_{12}$, $C^{13}\,_{13}$ for $m=-\frac{1}{2}$ leads to conclusion that
$F_i$ are polynomials of second order. Finally we obtain
 \be
 g=\frac{1}{M^2}\sum_{i=1}^4\frac{(dx^i)^2}{\prod_{j\neq i}(x^i-x^j)\sum_{k=0}^2 a_k(x^i)^k}, \label{met-1/2}
 \end{equation} 
where $a_k$ are constants.$\square$

\section{Conclusions}
We found general forms of binary 4-dimensional metrics (see (\ref{bin_metr}) and Table \ref{tab}) for which  
$C_{ijij}$ are the only non-vanishing components of the Weyl tensor.
All conformally flat metrics of case iv) has been found in explixit form, see (\ref{met0}), (\ref{met1/2}), 
(\ref{met-1}) and (\ref{met-1/2}). All of them  could be Riemannian, Lorentzian or neutral.
Moreover in generic case neither of three solutions (\ref{met1/2}), (\ref{met-1}), (\ref{met-1/2}) 
are flat even for $M=const$.

It is not difficult to construct an example of Lorentzian metric. Assume that 
polynomial $\sum_{i=0}^6 a_k(x^i)^6$ has exactly four real different roots $b_i$ and $a_6>0$. If the
 coordinates $(x^i)$ satisfy inequalities
 \be
    x^1>b_1>x^2>b_2>x^3>b_3>b_4>x_4
 \ee 
than metric (\ref{met1/2}) has signature ($+++-$). The Lorentzian metrics could be potentially useful in general
relativity where they represent spacetime.  
Metric (\ref{met1/2}) is flat if $M$ is a constant function and $a_5=a_6=0$. If in addition 
polynomial $a_0+a_1z+a_2z^2+a_3 z^3+a_4 z^4$ has four real different roots then $(x^i)$ constitute
elliptic coordiates in Euclidean flat space.

\section*{Acknowledgments}

This work is partially supported by the grant N N202 104838 of Ministry of Science and Higher 
Education of Poland.

\end{document}